\documentclass[prd,aps,showpacs,tightenlines,twocolumn,nofootinbib,nobibnotes]{revtex4}
\usepackage{amsmath}
\usepackage{graphicx}
\usepackage{color}
\usepackage{amssymb}
\usepackage{latexsym}

\newcommand{\ee}{e}
\newcommand{\sss}[1]{{\scriptscriptstyle{#1}}}
\newcommand{\order}[1]{\mathcal{O}\!\left(#1\right)}

\newcommand{\eps}[1]{\epsilon_{#1}}
\newcommand{\alp}[1]{\alpha_{#1}}
\newcommand{\del}[1]{\delta_{#1}}

\newcommand{\Mpl}{M_{_\mathrm{P}}}

\newcommand{\calH}{\mathcal{H}}
\newcommand{\calP}{\mathcal{P}}
\newcommand{\calPh}{\calP_h}
\newcommand{\calPz}{\calP_\zeta}
\newcommand{\uT}{\mathrm{T}}
\newcommand{\uS}{\mathrm{S}}
\newcommand{\ud}{\mathrm{d}}
\newcommand{\us}{\mathrm{s}}

\newcommand{\nS}{n_{\sss{\uS}}}
\newcommand{\nT}{n_{\sss{\uT}}}
\newcommand{\alphaT}{\alpha_{\sss{\uT}}}
\newcommand{\alphaS}{\alpha_{\sss{\uS}}}
\newcommand{\betaS}{\beta_{\sss{\uS}}}
\newcommand{\betaT}{\beta_{\sss{\uT}}}
\newcommand{\cs}{c_\us}
\newcommand{\kstar}{k_*}
\newcommand{\etastar}{\eta_*}
\newcommand{\Hstar}{H_*}
\newcommand{\epsstar}[1]{\eps{#1*}}

\newcommand{\alpstar}[1]{\alp{#1*}}
\newcommand{\Ng}{\tilde{N}}
\newcommand{\Hg}{\tilde{H}}
\newcommand{\Hgstar}{\Hg_*}
\newcommand{\calHg}{\tilde{\calH}}
\newcommand{\epssnd}[1]{\eps{#1\triangleright}}
\newcommand{\delsnd}[1]{\del{#1\triangleright}}
\newcommand{\Hsnd}{H_\triangleright}
\newcommand{\cssnd}{c_{\us\triangleright}}
\newcommand{\ksnd}{k_{\triangleright}}
\newcommand{\tausnd}{\tau_\triangleright}
\newcommand{\etasnd}{\eta_\triangleright}
\newcommand{\nsnd}{n_\triangleright}
\newcommand{\psnd}{p_\triangleright}
\newcommand{\qsnd}{q_\triangleright}
\newcommand{\minidiam}{\sss{\diamond}}
\newcommand{\epsdiam}[1]{\eps{#1\minidiam}}
\newcommand{\deldiam}[1]{\del{#1\minidiam}}
\newcommand{\Hdiam}{H_\minidiam}
\newcommand{\csdiam}{c_{\us\minidiam}}
\newcommand{\kdiam}{k_{\minidiam}}

\newcommand{\etadiam}{\eta_\minidiam}
\newcommand{\ndiam}{n_\minidiam}
\newcommand{\pdiam}{p_\minidiam}
\newcommand{\qdiam}{q_\minidiam}

\begin{document}

\title{Exact Mapping between Tensor and Most General Scalar Power Spectra}

\author{Jose Beltr\'an Jimenez} \email{jose.beltran@uclouvain.be}
\affiliation{Centre for Cosmology, Particle Physics and Phenomenology,
  Institute of Mathematics and Physics, Louvain University, 2 Chemin
  du Cyclotron, 1348 Louvain-la-Neuve, Belgium}

\author{Marcello Musso}
\email{marcello.musso@uclouvain.be}
\affiliation{Centre for Cosmology, Particle Physics and Phenomenology,
  Institute of Mathematics and Physics, Louvain University, 2 Chemin
  du Cyclotron, 1348 Louvain-la-Neuve, Belgium}
\author{Christophe Ringeval}
\email{christophe.ringeval@uclouvain.be}

\affiliation{Centre for Cosmology, Particle Physics and Phenomenology,
  Institute of Mathematics and Physics, Louvain University, 2 Chemin
  du Cyclotron, 1348 Louvain-la-Neuve, Belgium}
\date{\today}

\begin{abstract}
  We prove an exact relation between the tensor and the scalar
  primordial power spectra generated during inflation. Such a mapping
  considerably simplifies the derivation of any power spectra as they
  can be obtained from the study of the tensor modes only, which are
  much easier to solve. As an illustration, starting from the second
  order slow-roll tensor power spectrum, we derive in a few lines the
  next-to-next-to-leading order power spectrum of the comoving
  curvature perturbation in generalized single field inflation with a
  varying speed of sound.
\end{abstract}


\maketitle

\section{Introduction}
Cosmic inflation is currently considered to be the standard lore to
explain the origin of the Cosmic Microwave Background (CMB)
anisotropies and the large scale structures of our Universe. In
addition to solving the so-called ``problems'' of the standard
Friedmann-Lema\^{\i}tre-Robertson-Walker (FLRW) model, inflation makes
definite predictions for the cosmological perturbations in the
earliest times of the Universe's history~\cite{Starobinsky:1979ty,
  Starobinsky:1980te, Guth:1980zm, Linde:1981mu}. At linear order, it
predicts an almost scale invariant power spectrum for the comoving
curvature perturbation $\zeta$ in complete agreement with the spectral
index measured in the most recent CMB data~\cite{Bennett:2012fp,
  Hinshaw:2012fq, Hou:2012xq, Sievers:2013wk, Planck:2013kta,
  Ade:2013zuv}. Confronting the predictions of inflationary models
with increasingly more accurate cosmological data has pushed forward
various theoretical developments. Among them, the search for
non-Gaussianities has triggered interest in the calculation of higher
$n$-point functions for $\zeta$ which are expected to trace any
departures from a single slow-rolling field~\cite{Seery:2005wm,
  Chen:2006nt, Musso:2006pt, Suyama:2007bg, Huterer:2010en,
  Creminelli:2011rh}. This is particularly timely as the Planck
satellite has severely constrained the amount of possible
non-Gaussianities in the CMB data~\cite{Ade:2013ydc, Ade:2013xla}
while dramatically increasing the accuracy in the measurement of the
scalar spectral index. Reference~\cite{Ade:2013zuv} reports
$\nS=0.9603 \pm 0.0073$ using Planck temperature data complemented
with WMAP polarization~\cite{Bennett:2012fp}. Both of these results
suggest that admissible inflationary models cannot be too far from the
slow-roll single field inflation paradigm. Within this landscape of
models, the shape of the primordial power spectra is an observable of
choice to discriminate between various scenarios. In this respect
several complementary approaches have been proposed. Given a model of
inflation, it is always possible to exactly evaluate the power
spectrum using numerical
methods\footnote{\url{http://theory.physics.unige.ch/~ringeval/fieldinf.html}},
eventually complemented with Bayesian model comparison to determine
how well they suit in any cosmological data set~\cite{Martin:2006rs,
  Ringeval:2007am, Martin:2010hh, Easther:2011yq}.

A second method, which is the one we will be interested in, consists
in parametrizing the power spectra of a broad class of models using
the slow-roll expansion~\cite{Mukhanov:1985rz, Stewart:1993bc,
  Martin:1999wa, Schwarz:2001vv, Leach:2002ar, Schwarz:2004tz}. The
modern approach consists in defining an infinite hierarchy of
so-called Hubble flow functions~\cite{Hoffman:2000ue, Schwarz:2001vv}
(also simply referred to as the slow-roll parameters)
\begin{equation}
\eps{i+1} \equiv \dfrac{\ud \ln |\eps{i}|}{\ud N}, \qquad
\eps{1} \equiv -\dfrac{\ud \ln H}{\ud N}\,,
\label{eq:hubbleflow}
\end{equation}
where $H$ is the the Hubble parameter and $N\equiv \ln a$ with $a$ the
scale factor during inflation. By definition, the expansion of the
Universe is accelerated if $\eps{1}<1$ and the slow-roll approximation
relies on the extra-assumptions that $\eps{1} \ll 1$ and all the
$\eps{i}$ are of the same order of magnitude that we denote by
$\order{\eps{}}$. Provided this is verified, one can consistently
solve order by order the evolution equations for the cosmological
perturbations. After some appropriate field redefinitions, the
equations for the two polarization degrees of freedom $h_\lambda$ of
the tensor modes and for the scalar comoving curvature perturbation
$\zeta$ can all be written in Fourier space in terms of a
Mukhanov-Sasaki variable $v$ verifying~\cite{Garriga:1999vw}
\begin{equation}
v'' + \left(k^2 - \dfrac{z''}{z} \right) v = 0,
\label{eq:modevol}
\end{equation}
where a ``prime'' denotes differentiation with respect to an
appropriate time variable $\tau$, and $z$ is a suitable function of
$\tau$. For instance, in canonical single field inflation $\tau=\eta$
is the standard conformal time ($\ud \eta = \ud t/a$); for the tensor
modes $v(k,\eta) \equiv h_\lambda(k,\eta)\,z(\eta)$ with $z(\eta)
\equiv a(\eta)$~\cite{Starobinsky:1979ty}, while for the scalar mode
$v(k,\eta) \equiv \sqrt{2} \, \zeta(k,\eta) \, z(\eta)$ with $z(\eta)
\equiv a(\eta) \sqrt{\eps{1}(\eta)}$~\cite{Mukhanov:1990me}.

It is well known that this equation remains the same for
any single field model with the most generic quadratic action, such as
K-inflation~\cite{Silverstein:2003hf, Alishahiha:2004eh,
  Kinney:2007ag, Cheung:2007st, Langlois:2008qf, DeFelice:2011bh,
  Baumann:2011dt}. In that case, $\tau$ is a rescaled conformal time
defined by $\ud \tau = \cs(\eta) \ud \eta$, where $\cs$ stands for the
``sound speed'' associated with the scalar perturbations. The Mukhanov
variable still reads $v(k,\tau) \equiv \sqrt{2} \, \zeta(k,\tau) \,
z(\tau)$, with now $z(\tau) \equiv a(\tau)
\sqrt{\eps{1}(\tau)/\cs(\tau)}$.

The slow-roll approximation allows to consistently solve
Eq.~\eqref{eq:modevol}, at a given order of approximation. For
standard single field models, the first calculation of this kind was
done in Ref.~\cite{Starobinsky:1979ty} for the tensor modes and in
Refs.~\cite{Mukhanov:1985rz, Mukhanov:1988jd} for the scalar
modes. The next-to-leading order corrections were first derived in
Ref.~\cite{Stewart:1993bc}. The scalar mode solution was then
rederived and extended using the more general Green function method in
Ref.~\cite{Gong:2001he}. These results have been also recovered for
both scalar and tensor modes by using the Wentzel-Kramers-Brillouin
(WKB) approximation in Ref.~\cite{Martin:2002vn} and the uniform
approximation in Refs.~\cite{Habib:2002yi,
  Habib:2004kc}. Next-to-next-to-leading order corrections for the
scalar spectral index were first derived in Ref.~\cite{Gong:2001he}
while the expanded power spectrum at second order has been explicitly
derived in Ref.~\cite{Schwarz:2001vv} for the scalars and in
Ref.~\cite{Leach:2002ar} for the tensor modes, still using the Green
function method. These results have been recovered with an improved
WKB approximation in Refs.~\cite{Casadio:2004ru, Casadio:2005xv}. As
we will need its expression later on, the second order tensor power
spectrum obtained in Ref.~\cite{Leach:2002ar} reads
\begin{widetext}
\begin{equation}
\begin{aligned}
  \calPh & = \dfrac{2 \Hstar^2}{\pi^2 \Mpl^2} \left\{ 1 -
    2(1+C)\epsstar{1} +\left(\dfrac{\pi^2}{2}-3 + 2C +2C^2 \right)
    \epsstar{1}^2 + \left(\dfrac{\pi^2}{12} -2 -2C -C^2 \right)
    \epsstar{1} \epsstar{2} \right. \\
  & + \left. \left[ -2 \epsstar{1} + (2+4C) \epsstar{1}^2 -2(1+C)
      \epsstar{1} \epsstar{2} \right]
    \ln\left(\dfrac{k}{\kstar}\right) + \left(2 \epsstar{1}^2 -
      \epsstar{1}\epsstar{2} \right)
    \ln^2\left(\dfrac{k}{\kstar}\right) \right\},
\end{aligned}
\label{eq:powerh}
\end{equation}
\end{widetext}
where $\Mpl^2 = 8 \pi G$ is the reduced Planck mass and $C$ is a
constant equal to $C=\gamma + \ln 2 -2 \simeq -0.729637$.  For the
sake of clarity, let us emphasize that this expression simply comes
from the integration of Eq.\eqref{eq:modevol} by keeping all
\emph{functions} of order less than or equal to $\order{\eps{}^2}$, and
dropping the higher order ones. Moreover, the power spectrum has been
expanded around an unique pivot scale, $\kstar$, such that all
``star'' quantities are evaluated at the time $\etastar$ defined by
$\kstar \etastar = -1$. This last step is necessary in order to 
explicit the dependence on the wavenumber $k$. Let us notice that the
spectral index $\nT = \ud \ln \calPh/\ud \ln k|_{\kstar}$ and the
running $\alphaT \equiv \ud^2 \ln \calPh/\ud (\ln k)^2|_{\kstar}$ are
immediately obtained by expanding the logarithm of
Eq.~\eqref{eq:powerh}\footnote{At second order, it is also crucial to
  specify the pivot scale definition. In particular, the pivot of
  Refs.~\cite{Gong:2001he, Leach:2002ar} is set at $k = aH$ instead of
  $\kstar \etastar = -1$ here. This is the reason why some numerical
  coefficients in Eq.~\eqref{eq:powerh} are different in front of the
  second order terms.}. 

As one can check in the references mentioned earlier, the scalar
mode calculations are usually much more involved than those for the
tensors. The technical difficulties in solving the scalar equations
are exacerbated when one wants to apply these techniques to models 
for which the perturbations propagate with a
varying speed of sound $\cs(\eta)$. A first attempt at next-to-leading
order was performed in Refs.~\cite{Shandera:2006ax, Bean:2007hc,
  Peiris:2007gz} but their results were implicitly assuming a constant
$\cs$ leading to some missing terms in the first order corrections to
the power spectrum amplitude. The first consistent calculation for the
scalar modes at next-to-leading order in K-inflation was presented in
Ref.~\cite{Kinney:2007ag} with the Green functions and in
Ref.~\cite{Lorenz:2008et} by means of the uniform approximation. There
is also a non-trivial dependence in $\cs$ arising in the tensor power
spectrum due to the pivot shift between scalars and tensors. This has
been first discussed and derived in Refs.~\cite{Lorenz:2008et,
  Agarwal:2008ah}. Finally, next-to-next-to-leading order corrections
have only been derived very recently in Ref.~\cite{Martin:2013uma}
within the uniform approximation only.

As we have just summarized, all the integration techniques
performed so far have been independently applied to either the tensor
or the scalar modes, at a given order, and for a given class of single
field models. In this work, we derive an exact mapping between all
the power spectra by noticing that even though the integration methods
are different, they all start from the same functional form, namely
Eq.~\eqref{eq:modevol}. In the next section, we explicitly prove the
existence of such a transformation by introducing some generalized flow
functions, very similar to the usual ones of
Eq.~\eqref{eq:hubbleflow}. Then we apply our method to the Green
function integration approach and derive, for the first time and in a
few lines, the power spectrum of the curvature perturbation at second
order for K-inflation.

\section{Generalized Flow functions}

{}From Eq.~\eqref{eq:modevol}, one can reinterpret the function
$z(\tau)$ as being a generalized scale factor from which one could
define a generalized $\ee$-fold number $\Ng \equiv \ln z$ and a
generalized Hubble parameter $\Hg$ with its conformal analogue $\calHg
\equiv z \Hg$ such that
\begin{equation}
\calHg(\tau) \equiv \dfrac{z'}{z} = \dfrac{\ud \Ng}{\ud \tau}\,.
\label{eq:genhubble}
\end{equation}
As a result, one can construct an infinite hierarchy of generalized
flow functions $\alp{i}$, exactly as in Eq.~\eqref{eq:hubbleflow}, but
based on this rescaled Hubble parameter:
\begin{equation}
\alp{i+1} \equiv \dfrac{\ud \ln |\alp{i}|}{\ud \Ng}, \quad
\alp{1} \equiv -\dfrac{\ud \ln \Hg}{\ud \Ng}\,.
\label{eq:genflow}
\end{equation}
In terms of these generalized quantities, we know that, up to an
overall normalization accounting for the different quantum initial
conditions for scalars versus tensors (such as the number of
polarization states), the power spectrum of any quantity at second
order using the Green function method must be given by
Eq.~\eqref{eq:powerh} with the replacement
\begin{equation}
  \epsstar{i} \rightarrow \alpstar{i}, \qquad \Hstar \rightarrow
  \Hgstar.
\label{eq:map}
\end{equation}
Even though this might seem a trivial remark, as it stems from 
well-known field redefinitions in the quadratic action for the 
perturbations, up to our knowledge this property has never been used 
before to actually solve the equations of motion and compute the 
relevant observables with the accuracy achieved in this paper. 
These generalized quantities are the only ones that
we can measure by detecting the amplitude and spectral index of the
scalar perturbations alone. From a purely effective point of view, it
has been noticed that the interplay of $\eps{1}$ and $\cs$ can lead to
an exactly scale invariant power spectrum also for finite values of
$\eps{1}$, at the cost of breaking the scale invariance of the tensor
modes~\cite{Khoury:2008wj} and increasing the amount of 
non-Gaussianity~\cite{Baumann:2011dt}. Even more radically, scale 
invariant scalar perturbations can also be obtained in 
non-inflationary backgrounds if one relies on other mechanisms to 
solve the horizon and flatness problem~\cite{ArmendarizPicon:2003ht, 
Khoury:2011ii}.

Assuming that $\cs$ is a free function, from
Eqs.~\eqref{eq:genhubble} and \eqref{eq:genflow} we have
\begin{equation}
\begin{aligned}
  \Ng & = N + \dfrac{1}{2} \left( \ln\eps{1} - \ln \cs \right),\\
  \Hg  & = \dfrac{H}{\sqrt{\eps{1} \cs}} \left(1 + \dfrac{\eps{2}
      + \del{1}}{2} \right),
\end{aligned}
\label{eq:mapNH}
\end{equation}
where we have defined the usual sound flow hierarchy~\cite{Lorenz:2008et}
\begin{equation}
\del{i+1} \equiv \dfrac{\ud \ln |\del{i}|}{\ud N}, \qquad
\del{1} \equiv -\dfrac{\ud \ln \cs}{\ud N}\,.
\end{equation}
It is important to notice that Eq.~\eqref{eq:mapNH} contains exact
functional relations between the usual Hubble and sound flow functions
and the generalized ones. As such, they can be used in \emph{any}
approximation schemes or expansions. They are also complete as they
fix by recurrence the mapping of the full hierarchy. For instance, using
$\ud N/\ud \Ng = [1+(\eps{2} + \del{1})/2]^{-1}$
the exact functional relations for the first two generalized flow
functions $\alp{1}$ and $\alp{2}$ are
\begin{widetext}
\begin{equation}
\begin{aligned}
  \alp{1} & = \dfrac{1}{ \left(1 + \dfrac{\eps{2} + \del{1}}{2}
    \right)^2 } \left(\eps{1} + \dfrac{1}{2} \eps{2} - \dfrac{1}{2}
      \del{1} + \dfrac{1}{2} \eps{1} \eps{2} + \dfrac{1}{4} \eps{2}^2
      - \dfrac{1}{2} \eps{2} \eps{3} + \dfrac{1}{2} \eps{1} \del{1} -
      \dfrac{1}{2} \del{1} \del{2} - \dfrac{1}{4}\del{1}^2 \right) , \\
    \alp{2} & = \dfrac{1}{1 + \dfrac{\eps{2} + \del{1}}{2}}
    \left(-\dfrac{\eps{2} \eps{3} + \del{1} \del{2}}{1 +
        \dfrac{\eps{2}
          + \del{1}}{2}} \right. \\
    & + \left. \dfrac{ 2 \eps{1} \eps{2} + \eps{2} \eps{3} - \del{1}
        \del{2} + \eps{1} \eps{2}^2 + \eps{1} \eps{2} \eps{3} +
        \eps{2}^2 \eps{3} - \eps{2} \eps{3}^2 - \eps{2} \eps{3}
        \eps{4} + \eps{1}\eps{2}\del{1} + \eps{1}\del{1}\del{2} -
        \del{1}\del{2}^2 - \del{1}\del{2}\del{3} - \del{1}^2\del{2}
      }{2 \eps{1} + \eps{2} - \del{1} + \eps{1}\eps{2} + \dfrac{1}{2}
        \eps{2}^2 - \eps{2}\eps{3} + \eps{1}\del{1} - \del{1}\del{2} -
        \dfrac{1}{2} \del{1}^2} \right).
\end{aligned}
\label{eq:mapalp}
\end{equation}
\end{widetext}
Standard single field inflation is recovered by plugging $\cs=1$ and
$\del{i}=0$ in Eqs.~\eqref{eq:mapNH} and \eqref{eq:mapalp}. In the
following, motivated by the Planck results, we adopt a conservative
approach by assuming $\cs$ is a free but slowly varying function such
that $\delta_i \sim \order{\eps{}}$.

\section{Power spectra with varying speed of sound}

The above mapping can now be applied to straightforwardly derive the
second order power spectrum for the comoving curvature perturbation in
generalized single field models with varying speed of sound. Plugging
Eqs.~\eqref{eq:mapNH} and \eqref{eq:mapalp} into
Eq.~\eqref{eq:powerh}, Taylor expanding everything at second order in
the $\eps{i}$ and $\del{i}$ parameters yields the desired scalar
spectrum. One should also not forget to divide the result by the
well-known factor $16$ which accounts for the different normalization
of the scalar action with respect to the one for tensors, arising from
the number of graviton polarization states~\cite{Mukhanov:1990me}. 
One finally gets
\begin{widetext}
\begin{equation}
\begin{aligned}
  \calPz & = \dfrac{\Hsnd^2}{8 \pi^2 \Mpl^2 \epssnd{1} \cssnd} \left\{
    1 -2(1+C)\epssnd{1} - C \epssnd{2} + (2+C) \delsnd{1} +
    \left(\dfrac{\pi^2}{2} -3 + 2C + 2C^2 \right) \epssnd{1}^2 +
    \left( \dfrac{7 \pi^2}{12} -6 -C+C^2\right) \epssnd{1} \epssnd{2}
  \right. \\ & + \left. \left(\dfrac{\pi^2}{8} -1 +\dfrac{C^2}{2}
    \right) \epssnd{2}^2 +\left(\dfrac{\pi^2}{24} - \dfrac{C^2}{2}
    \right) \epssnd{2} \epssnd{3} + \left(\dfrac{\pi^2}{8}
      \boldsymbol{+ \nsnd} +C + \dfrac{C^2}{2} \right) \delsnd{1}^2 +
    \left(-\dfrac{\pi^2}{24}
      +2 + 2C + \dfrac{C^2}{2} \right) \delsnd{1}\delsnd{2} \right. \\
  & + \left.  \left(-\dfrac{\pi^2}{2} \boldsymbol {+ \psnd} -3C -2C^2
    \right)\delsnd{1}\epssnd{1} + \left(-\dfrac{\pi^2}{4}
      \boldsymbol{+ \qsnd} -C
      -C^2\right) \delsnd{1} \epssnd{2} \right. \\& + \left. \bigg[-2
    \epssnd{1} - \epssnd{2} + \delsnd{1} +(2+4C) \epssnd{1}^2 + (-1 +
    2C) \epssnd{1} \epssnd{2} + C \epssnd{2}^2 -C \epssnd{2}
    \epssnd{3} + (1+C) \delsnd{1}^2 + (2+C) \delsnd{1} \delsnd{2}
  \right. \\& - \left. (3+4C) \delsnd{1} \epssnd{1} -(1+2C)\delsnd{1}
    \epssnd{2}\bigg] \ln\left(\dfrac{k}{\ksnd}\right) \right.  \\ & +
  \left. \bigg[2 \epssnd{1}^2 + \epssnd{1} \epssnd{2} + \dfrac{1}{2}
    \epssnd{2}^2 - \dfrac{1}{2} \epssnd{2} \epssnd{3} + \dfrac{1}{2}
    \delsnd{1}^2 + \dfrac{1}{2} \delsnd{1} \delsnd{2} - 2 \delsnd{1}
    \epssnd{1} - \delsnd{1} \epssnd{2} \bigg] \ln^2
    \left(\dfrac{k}{\ksnd}\right) \right\},
\end{aligned}
\label{eq:powerzsnd}
\end{equation}
\end{widetext}
where the three constants $\nsnd$, $\psnd$ and $\qsnd$ read
\begin{equation}
\nsnd = 0, \qquad \psnd = 2, \qquad \qsnd = 2.
\end{equation}
The new index ``$\triangleright$'' is different from ``$*$'' of
Eq.~(\ref{eq:powerh}). Indeed, it is important to understand that the
mapping method automatically induces a transformation on the pivot
definition. Starting from the tensor mode pivot of
Eq.~\eqref{eq:powerh} defined at $\kstar \etastar = -1$, we get the
scalar pivot defined in the same way but with the transformed
quantities, i.e. at $\ksnd \tausnd = -1$. As a result,
Eq.~\eqref{eq:powerzsnd} is expressed in terms of quantities evaluated
at the time $\etasnd$ such that
\begin{equation}
\ksnd \int_{\etasnd}^0 \cs(\eta) \ud \eta = -1.
\label{eq:pivsnd}
\end{equation}
Within standard single field models, i.e. those having $\cs(\eta)=1$,
there is no difference between the two pivots and $\etasnd = \etastar$
(at the same observable pivot mode $\ksnd=\kstar$).

The expression of Eq.~\eqref{eq:powerzsnd} has never been derived
before, but we can make some cross-checks with other approximation
methods. First of all, setting $\cssnd=1$ and $\delsnd{i}=0$, we
recover exactly the same expression as in Refs.~\cite{Gong:2001he,
  Leach:2002ar}, once the pivot has been switched from $k=aH$ to ours,
i.e. $\ksnd \etasnd = -1$ for $\cssnd=1$ (see the discussion
above). In the general case, Ref.~\cite{Wei:2004xx} claims to have
performed such a second order expansion using Green functions but
their derivation does not include the pivot expansion and it is
assumed that $\cs \simeq 1$, which makes it hardly comparable with our
result. On the other hand, as for the tensor modes, the spectral index
$\nS-1$ and the running $\alphaS$ at second order can be immediately
read out from the logarithm of Eq.~\eqref{eq:powerzsnd}. 
These quantities have already been derived in the literature, as for
instance in Refs.~\cite{Kinney:2007ag, Lorenz:2008et}, using the 
trick described in Ref.~\cite{Schwarz:2004tz}, which allows to derive
spectral index and running at second order from the power spectrum at
first order. Our results match both expressions, and we do not repeat
them here. However, applying the method of Ref.~\cite{Schwarz:2004tz}
to our results gives now the spectral index and running at third 
order (see Appendix). Finally, Ref.~\cite{Martin:2013uma} has recently
derived the very same power spectrum, at the pivot scale $\kdiam
\etadiam \csdiam = -1$, by using the uniform approximation to
directly solve the scalar equation of motion. Up to the well-known
differences between the Green function method and the WKB/uniform
approximations, Eq.~\eqref{eq:powerzsnd} is compatible with this
reference after one has performed the change of pivot described
below. A last check we have performed is to apply the mapping
technique within the uniform approximation scheme. Starting from the
tensor power spectrum at second order given in
Ref.~\cite{Martin:2013uma}, using Eqs.~\eqref{eq:mapNH} and
\eqref{eq:mapalp}, we have reproduced the second order power spectrum
in the uniform approximation derived in that reference.

To be complete, we would like to express the power spectrum at the
more widespread pivot $\etadiam$ defined by
\begin{equation}
\kdiam \etadiam \csdiam = -1.
\end{equation}
Changing from one pivot to the other is a straightforward, but
lengthy, calculation that requires performing slow-roll expansions
for all terms of Eq.~\eqref{eq:powerzsnd}. Details on such a
transformation can be found in Ref.~\cite{Lorenz:2008et} and we simply
here report the result. One gets exactly the same expression as
Eq.~\eqref{eq:powerzsnd} but with three different numerical
coefficients for $n$, $p$ and $q$ given by:
\begin{equation}
\ndiam =-1, \qquad \pdiam=4, \qquad \qdiam=3.
\end{equation}
These numbers are only involved in the overall amplitude, i.e. not in
front of any $k$-dependent terms and therefore this change of pivot
does not affect the spectral index and running at second order (it
does at third order). With this new pivot, we have checked that the
numerical values of all multiplying coefficients is within a few
percents to those given by the uniform approximation of
Ref.~\cite{Martin:2013uma}, even though they are defined from
different combinations of irrational numbers and stem from a complete
different approach to solve the equations of motion.

{}From the data analysis point of view, one should simultaneously use
both the scalar and tensor power spectra. In particular, this allows
to measure, or bound, the tensor-to-scalar ratio. In order to get
meaningful results it is however crucial to evaluate them at the same
pivot. From Eq.\eqref{eq:powerh}, moving the pivot from $\kstar
\etastar = -1$ to $\kdiam \etadiam \csdiam = -1$, one gets
\begin{widetext}
\begin{equation}
\begin{aligned}
  \calPh & = \dfrac{2 \Hdiam^2}{\pi^2 \Mpl^2 } \left\{1 - 2 (1 + C -\ln
  \csdiam) \epsdiam{1} + \left[\dfrac{\pi^2}{2} - 3 +2C +2C^2 -
    (2+4C)\ln \csdiam + 2 \ln^2\csdiam \right] \epsdiam{1}^2 \right. \\
  & \left. + \left[ \dfrac{\pi^2}{12} -2 -2C -C^2 + 2(1+C)\ln\csdiam -\ln^2
    \csdiam \right] \epsdiam{1} \epsdiam{2} \right. \\ &
  + \left. \left[-2 \epsdiam{1} +
    (2+4C-4\ln \csdiam) \epsdiam{1}^2 -2(1+C-\ln\csdiam) \epsdiam{1}
    \epsdiam{2} \right] \ln\left( \dfrac{k}{\kdiam} \right) + \left(2
  \epsdiam{1}^2 - \epsdiam{1} \epsdiam{2} \right) \ln^2
  \left(\dfrac{k}{\kdiam} \right) \right\},
\end{aligned}
\label{eq:powerhdiam}
\end{equation}
\end{widetext}
which now explicitly depends on $\csdiam$.

\section{Conclusion}

We have derived a simple transformation that is summarized by
Eqs.~\eqref{eq:genflow}, \eqref{eq:map} and \eqref{eq:mapNH} which
allows to map the tensor mode perturbations into the scalar ones. This
transformation being exact, it can be used at any order of a flow
expansion and within any approximation schemes to integrate the mode
equation. We have illustrated its usefulness by deriving for the first
time the second order power spectrum for the comoving curvature
perturbation using the Green function method and for generalized
single field inflation models having a varying speed of sound. 

It is important to stress that since $\order{\epsilon} =
\order{\nS-1}$, taking the Planck results quoted in the introduction,
one has $\order{\epsilon^2} \simeq 10^{-3}$. This number is of
comparable amplitude with the measurement accuracy of the spectral index
and shows that the Planck data are already sensitive to the second
order corrections. From a Bayesian data analysis point of view, it
means that even if the second order terms cannot yet be measured, they
should be included in the data analysis and marginalized over to allow
for a robust determination of the $\eps{i}$ at first order. Let us
also mention that these terms, and eventually the third order ones,
will be crucial in the context of 21-cm cosmology~\cite{Lewis:2007kz,
  Mao:2008ug, Adshead:2010mc, Clesse:2012th}. At last, and as it is
discussed in Ref.~\cite{Kuroyanagi:2011iw}, direct detection of
primordial gravitational waves requires higher order corrections
to be included in the tensor power spectrum because the observable
wave numbers are quite different from the ones the CMB is sensitive
to. For all these reasons, we give in the Appendix the third order
expression of the spectral index, its running and the running of the
running for both the tensor and scalar primordial power spectra.

Finally, as the calculations involving the tensor modes are far easier
than those involving the scalars, our approach opens the feasibility
window for higher order expansions. In principle, our mapping can also
be used directly for the perturbed variables and this could also
simplify the derivation of higher $n$-point functions involved in the
calculations of non-Gaussianities.

\appendix*
\section{Spectral index and runnings}

In this appendix, for completeness, we give the spectral index, the
running and the running of the running up to third order in slow-roll
parameters for both scalar and tensor perturbations. The expressions
for the scalar perturbations are
\begin{widetext}
\begin{align}
\nS-1 &
=-(2\epsdiam{1}+\epsdiam{2}-\deldiam{1})-2\epsdiam{1}^2-(3+2C)\epsdiam{1}\epsdiam{2}-C\epsdiam{2}\epsdiam{3}+3\deldiam{1}\epsdiam{1}+\deldiam{1}\epsdiam{2}-\deldiam{1}^2+(2+C)\deldiam{1}\deldiam{2}
\displaybreak[0]
\nonumber\\
&-2\epsdiam{1}^3-(15+6C-\pi^2)\epsdiam{1}^2\epsdiam{2}+5\deldiam{1}\epsdiam{1}^2-\left(7+3C+C^2-\frac{7\pi^2}{12}\right)\epsdiam{1}\epsdiam{2}^2-\left(6+4C+C^2-\frac{7\pi^2}{12}\right)\epsdiam{1}\epsdiam{2}\epsdiam{3}
\displaybreak[0]
\nonumber\\
&+\left(13+5C-\frac{\pi^2}{2}\right)\deldiam{1}\epsdiam{1}\epsdiam{2}-4\deldiam{1}^2\epsdiam{1}+\left(10+4C-\frac{\pi^2}{2}\right)\deldiam{1}\deldiam{2}\epsdiam{1}-\left(2-\frac{\pi^2}{4}\right)\epsdiam{2}^2\epsdiam{3}-\left(\frac{C^2}{2}-\frac{\pi^2}{24}\right)\epsdiam{2}\epsdiam{3}^2
\displaybreak[0]
\nonumber\\
&+\left(3+2C-\frac{\pi^2}{4}\right)\deldiam{1}\epsdiam{2}\epsdiam{3}-\left(\frac{C^2}{2}-\frac{\pi^2}{24}\right)\epsdiam{2}\epsdiam{3}\epsdiam{4}-\deldiam{1}^2\epsdiam{2}+\left(3+C-\frac{\pi^2}{4}\right)\deldiam{1}\deldiam{2}\epsdiam{2}+\deldiam{1}^3
\displaybreak[0]
\nonumber\\
&-\left(8+3C-\frac{\pi^2}{4}\right)\deldiam{1}^2\deldiam{2}
+\left(2+2C+\frac{C^2}{2}-\frac{\pi^2}{24}\right)\deldiam{1}\deldiam{2}^2+\left(2+2C+\frac{C^2}{2}-\frac{\pi^2}{24}\right)\deldiam{1}\deldiam{2}\deldiam{3},
\end{align}
\end{widetext}
\begin{equation}
\begin{aligned}
\alphaS & =-2\epsdiam{1}\epsdiam{2}-\epsdiam{2}\epsdiam{3}+\deldiam{1}\deldiam{2}
-6\epsdiam{1}^2\epsdiam{2} \\
&
-(3+2C)\epsdiam{1}\epsdiam{2}^2-2(2+C)\epsdiam{1}\epsdiam{2}\epsdiam{3}
+ 5\deldiam{1}\epsdiam{1}\epsdiam{2} \\
&+ 4\deldiam{1}\deldiam{2}\epsdiam{1}
-C\eps{2}\eps{3}^2-C\epsdiam{2}\epsdiam{3}\epsdiam{4}+2\deldiam{1}\epsdiam{2}\epsdiam{3}
\\ & +\deldiam{1}\deldiam{2}\epsdiam{2}
-3\deldiam{1}^2\deldiam{2}+(2+C)\deldiam{1}\deldiam{2}(\deldiam{2}+\deldiam{3}),
\end{aligned}
\end{equation}
\begin{equation}
\begin{aligned}
\betaS & =
-2\epsdiam{1}\epsdiam{2}^2-2\epsdiam{1}\epsdiam{2}\epsdiam{3} -
\epsdiam{2}\epsdiam{3}^2-\epsdiam{2}\epsdiam{3}\epsdiam{4} \\
& + \deldiam{1}\deldiam{2}\deldiam{3} + \deldiam{1} \deldiam{2}^2.
\end{aligned}
\end{equation}

The spectral index of the tensor mode power spectrum at third order
reads
\begin{equation}
\begin{aligned}
\nT & =-2\epsdiam{1}-2\epsdiam{1}^2-2(1+C-\ln\csdiam)\epsdiam{1}\epsdiam{2}
-2\epsdiam{1}^3  \\ & - (14+6C-\pi^2-6\ln\csdiam)\epsdiam{1}^2\epsdiam{2} \\ & 
-\left[2+2C+C^2-\frac{\pi^2}{12}-2(1+C)\ln\csdiam+\ln^2\csdiam\right]
\\ & \times \epsdiam{1}\epsdiam{2}(\epsdiam{2}+\epsdiam{3}),\\
\end{aligned}
\end{equation}
while the running is given by
\begin{equation}
\begin{aligned}
\alphaT & =-2\epsdiam{1}\epsdiam{2} - 6\epsdiam{1}^2\epsdiam{2} -
2(1+C-\ln\csdiam)\epsdiam{1}\epsdiam{2}^2 \\ & -2(1+C-\ln\csdiam)\epsdiam{1}\epsdiam{2}
\epsdiam{3},
\end{aligned}
\end{equation}
and the running of the running is
\begin{equation}
\begin{aligned}
\betaT & =-2\epsdiam{1}\epsdiam{2}(\epsdiam{2}+\epsdiam{3}).
\end{aligned}
\end{equation}

Finally, the tensor-to-scalar ratio, up to third order, is given by
\begin{widetext}
\begin{align}
 r & = 16\epsdiam{1}\csdiam\left\{1-(2+C)\deldiam{1}+C\epsdiam{2}+2\epsdiam{1}\ln\csdiam+\left(5+3C+\frac{C^2}{2}-\frac{\pi^2}{8}\right)\deldiam{1}^2-\left(2+2C+\frac{C^2}{2}-\frac{\pi^2}{24}\right)\deldiam{1}\deldiam{2}\right.\nonumber\\
 &-\left(3+3C+C^2-\frac{\pi^2}{4}\right)\deldiam{1}\epsdiam{2}+\left(1+\frac{C^2}{2}-\frac{\pi^2}{8}\right)\epsdiam{2}^2+\left(\frac{C^2}{2}-\frac{\pi^2}{24}\right)\epsdiam{2}\epsdiam{3}+2(1+\ln\csdiam)\ln\csdiam\epsdiam{1}^2\nonumber\\
 &\left.-\left[8+3C-\frac{\pi^2}{2}+2(2+C)\ln\csdiam\right]\deldiam{1}\epsdiam{1}+\left[4+C-\frac{\pi^2}{2}+2(1+2C)\ln\csdiam-\ln^2\csdiam\right]\epsdiam{1}\epsdiam{2}\right\}.
\end{align}
\end{widetext}

\begin{acknowledgments}
  This work is supported by the ESA Belgian Federal PRODEX Grant
  No.~4000103071 and the Wallonia-Brussels Federation Grant ARC
  No.~11/15-040.
\end{acknowledgments}

\bibliography{references}
\end{document}